# Star luminosity function as an age indicator for the Dwarf Spheroidal Leo I

**F. Caputo[1] V. Castellani[2,3] and S. Degl'Innocenti[4,5]**

[1] Istituto di Astrofisica Spaziale, CNR, C.P. 67, I-0044 Frascati, Italy
[2] Dipartimento di Fisica, Università di Pisa, Piazza Torricelli 3, I-56100 Pisa, Italy
[3] Osservatorio Astronomico Collurania, I-64100 Teramo, Italy
[4] Max-Planck Institut for Astrophysics, K. Schwarzschild Str. 1, 85740 Garching bei Munchen, Germany
[5] INFN Sezione di Ferrara, I-44100 Ferrara, Italy

**Abstract.** Star luminosity function, already recognized as an age indicator for old galactic globular clusters, can be used to contrains the age of younger stellar systems like the nearby dwarfs spheroidal Leo I. We compare the observed luminosity function of Leo I, presented by Lee et al. 1993, with theoretical expectations for three selected ages, 1, 1.5 and 2 billion years, deriving an age of about 1.5 Gyr. This result does not appear critically affected by assumptions about the cluster distance modulus or the Initial Mass Function.

## 1. Introduction

Even if not very popular nor very used, star luminosity function has been proved to be a powerful tool to gain information about the age of stellar systems as old as galactic globular clusters (see, e.g., Paczinsky 1984, Chieffi & Gratton 1986). In this paper we will show that such a procedure can be usefully extended to constrain the age of relatively younger stellar systems like the nearby dwarf spheroidal Leo I.

Observational data recently presented for Leo I by Lee et al. (1993: thereinafter L93) have already stimulated theoretical efforts to investigate the evolution of stars in similar not-too-old metal poor systems (Castellani & Degl' Innocenti 1995, Caputo & Degl' Innocenti 1995). According to these theoretical evaluations, it has been shown that the observed distribution of He burning stars in that galaxy suggests that Leo I should be even younger of the 3 billion years estimated by L93 (Caputo et al. 1995: Paper I). Such a suggestion will be now reinforced through the analysis of the luminosity function of H-burning stars, as presented by L93, allowing a closer determination of the cluster age.

*Send offprint requests to*: S. Degl'Innocenti, Dipartimento di Fisica Universita' di Pisa, Piazza Torricelli 3, I-56100 Pisa, Italy

**Fig. 1.** Visual magnitude of the tip of the RG branch as a function of the age for the two labelled metallicities and for an original helium abundance Y=0.23

## 2. Luminosity function for H-burning stars

To prepare the theoretical instrument, let us assume Z=0.0004 as a suitable choice for the cluster metallicity (see Paper I). To fit theory with observation we will follow the procedure adopted by Lee et al. assuming the tip of the red giant branch at V= 19.65. By matching this value with the tip of theoretical branches one finds the cluster distance modulus to be applied to shift theoretical minosity is depending not only on metallicity but also on the cluster age (Fig. 1), in this way one finds DM=21.97, 22.11 and 22.14 for the ages t= 1.0, 1.5 and 2.0 Gyr, respectively. Fig. 2 shows the cluster isochrones for the three selected ages 1, 1.5 and 2 billion year and for the various results about the cluster distance modulus.

**Fig. 2.** Cluster isochrones in the color-magnitude diagram for the three selected ages 1,1.5 and 2 Gyr (see text). The original helium abundance is Y=0.23 and the metallicity is Z=0.0004.

Fig.3 discloses the expected luminosity function (LF) of H-burning stars along the 1 Gyr isochrone for the labeled assumptions about the exponent of the Initial Mass function (IMF). Starting from the luminous end, one finds that the regular increase of the LF going down along the RG branch is first interrupted by the so called RG bump, with a final abrupt increase when reaching the basis of the branch. As an important point one finds that the LF appears honestly independent of the assumption about the IMF down, about, the luminosity of the cluster turn off. Thus the distribution of stars above the TO is a function of the cluster age only.

Fig. 4 shows the luminosity function of stars in Leo I as presented by L93 As already recognized by the above quoted authors, the evidence for the more luminous maximum in the distribution has to be attributed to the contribution of He burning stars, whereas the second fainter maximum is a "signal" of the approach to the turn off region. To allow the comparison with the observed luminosity function, the same fig. 4 shows the expected theoretical distributions of stars along the isochrones of Fig.2, as computed assuming a flat IMF.Theoretical luminosity functions have been normalized to the observed number of stars in the upper portion of the giant branch, that is for a luminosity higher than 20.5 magnitudes.

**Fig. 3.** Theoretical luminosity functions of H-burning stars along the 1 Gyr isochrone for the labeled assumptions about the exponent of the initial mass function

Inspection of Fig.4 easily discloses the required indication on the age. As a first point the figure shows that the maximum at about V= 23.5 should be regarded as an artifact of the photometry, the real distribution remaining at very large values down to very fainter magnitudes (see Fig.2) unless a very peculiar and rather unrealistic distribution of the IMF would be assumed. This is not a puzzling occurrence, since the same Lee et al. advised that the photometry at the larger magnitudes is far from being complete. However, in spite of such a lack of completeness, Fig 4 shows that the observed increase of counts around V=23 can be only understood if the cluster age is in the range of about 1.5 Gyr, since a shift in ages of 0.5 Gyr would shift the sudden increase of the LF at magnitudes hardly compatible with the observed distribution.

**Fig. 4.** Theoretical luminosity functions for Z=0.0004 and three different ages 1, 1.5 and 2 Gyr as computed assuming a flat IMF compared with the observational luminosity function presented by Lee et al. (1993).

One may notice that this result appears rather unaffected by honest variations either in the distance modulus

As a result, the comparison made in Fig. 4 tell us that, if Z=0.0004 and the tip of the RG branch is around V= 19.65, one cannot escape the conclusion that Leo I is as young as, about, 1.5 Gyr,(i.e., much younger than well known old galactic clusters like NGC188) with stars around 1.5 $M_\odot$ in the advanced evolutionary phases. As a further test of the consistency of observational data with theory, from evolutionary results one expects for the ratio between lifetimes of He burning to red giant stars a value of the order of 1.2, well consistent with the distribution reported in Fig.4. One could finally tempted to read in the observed LF a suggestion for the RG clump around V= 21, a suggestion which needs to be possibly confirmed by further observations.

## References

Caputo F., Castellani V. & Degl' Innocenti S. 1995, A&A (in publication)

Caputo F. & Degl' Innocenti S. 1995, A&A 298, 833

Castellani V. & Degl' Innocenti S. 1995, A&A 298, 827

Chieffi A. & Gratton G. 1986, Mem. Soc. Astron.It. 57, 398.

Lee M.G., Freedman W., Mateo M. & Thompson I. 1993, AJ 106,1420.

Paczinsky B. 1984, ApJ 284,670.

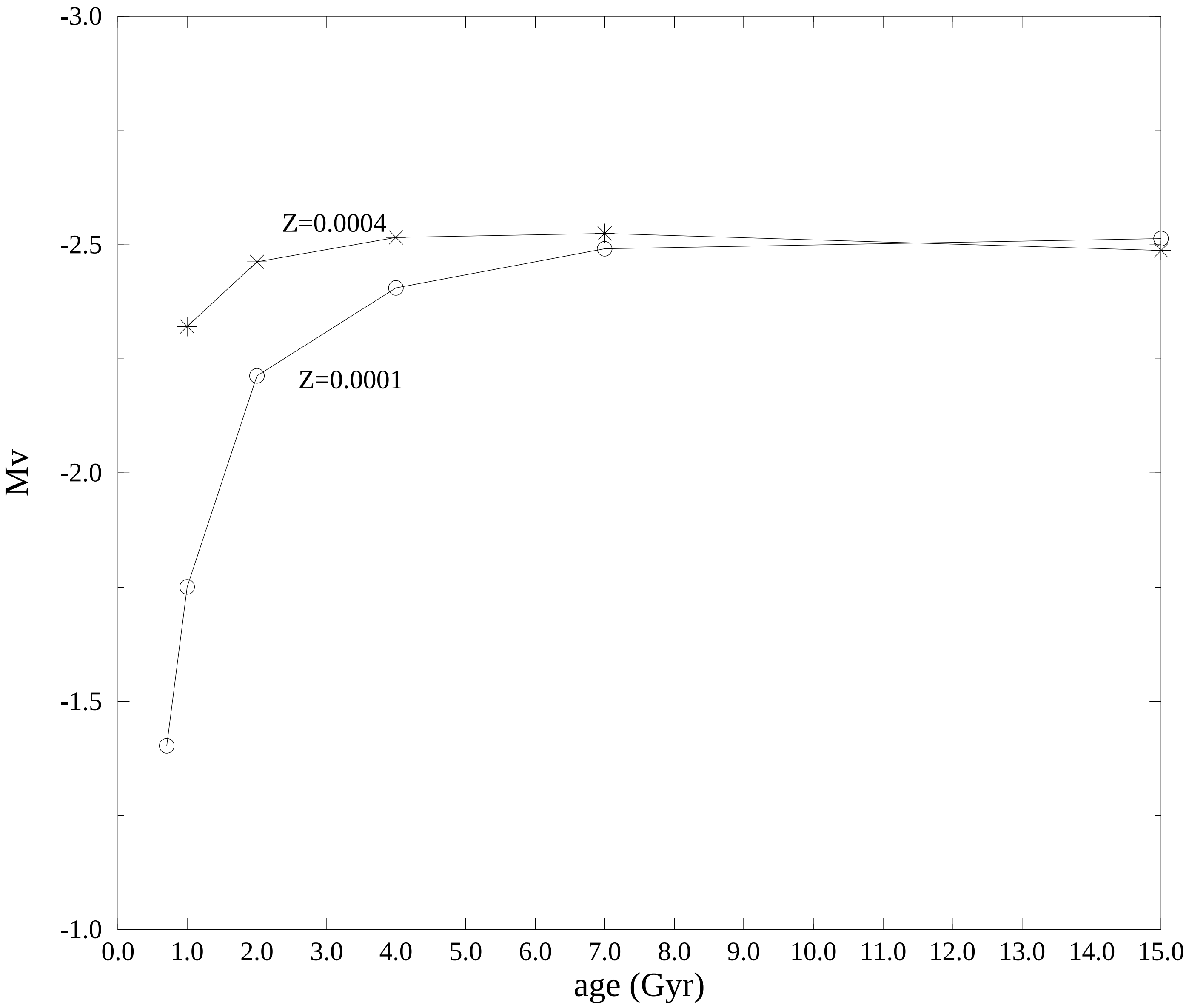

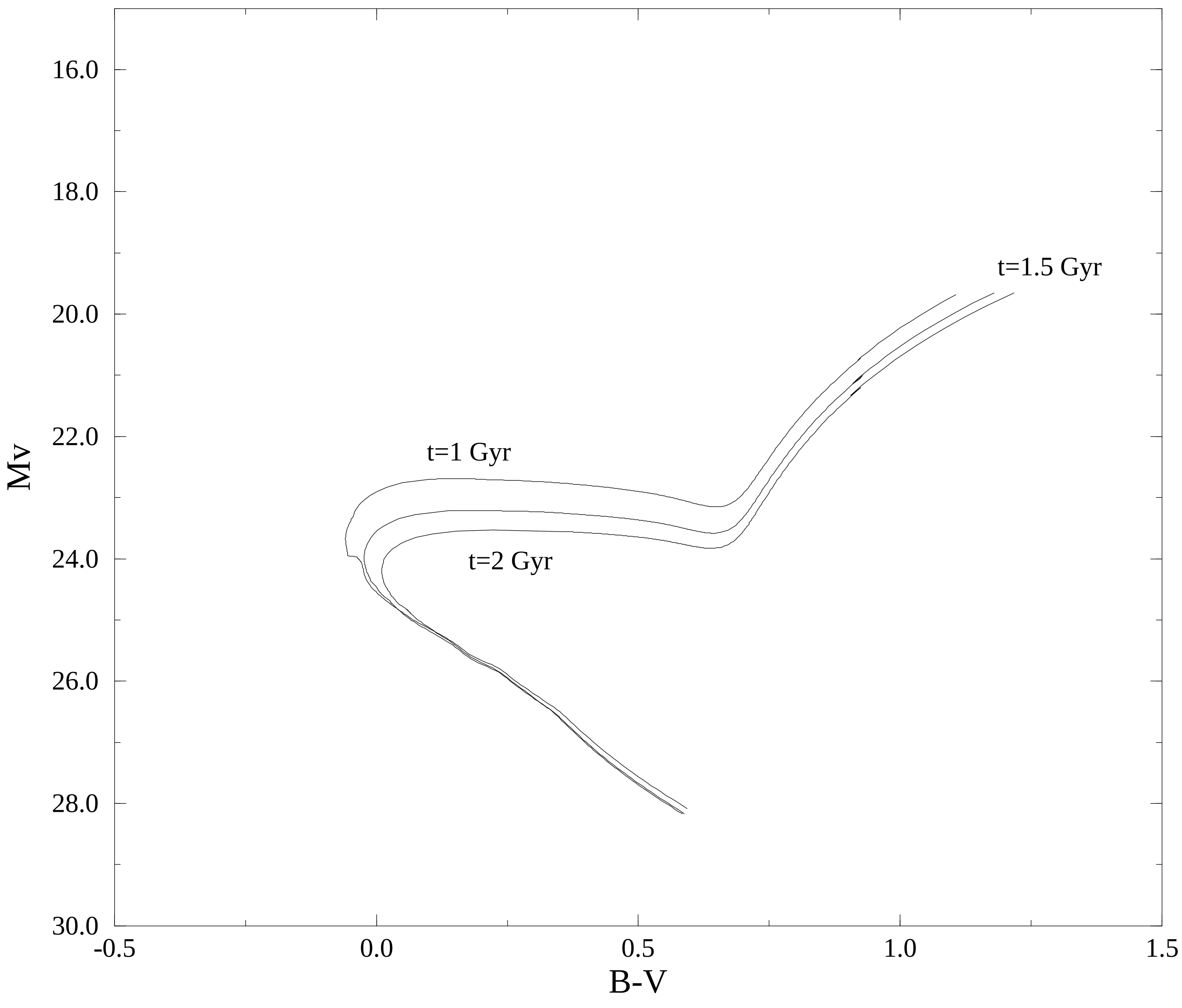

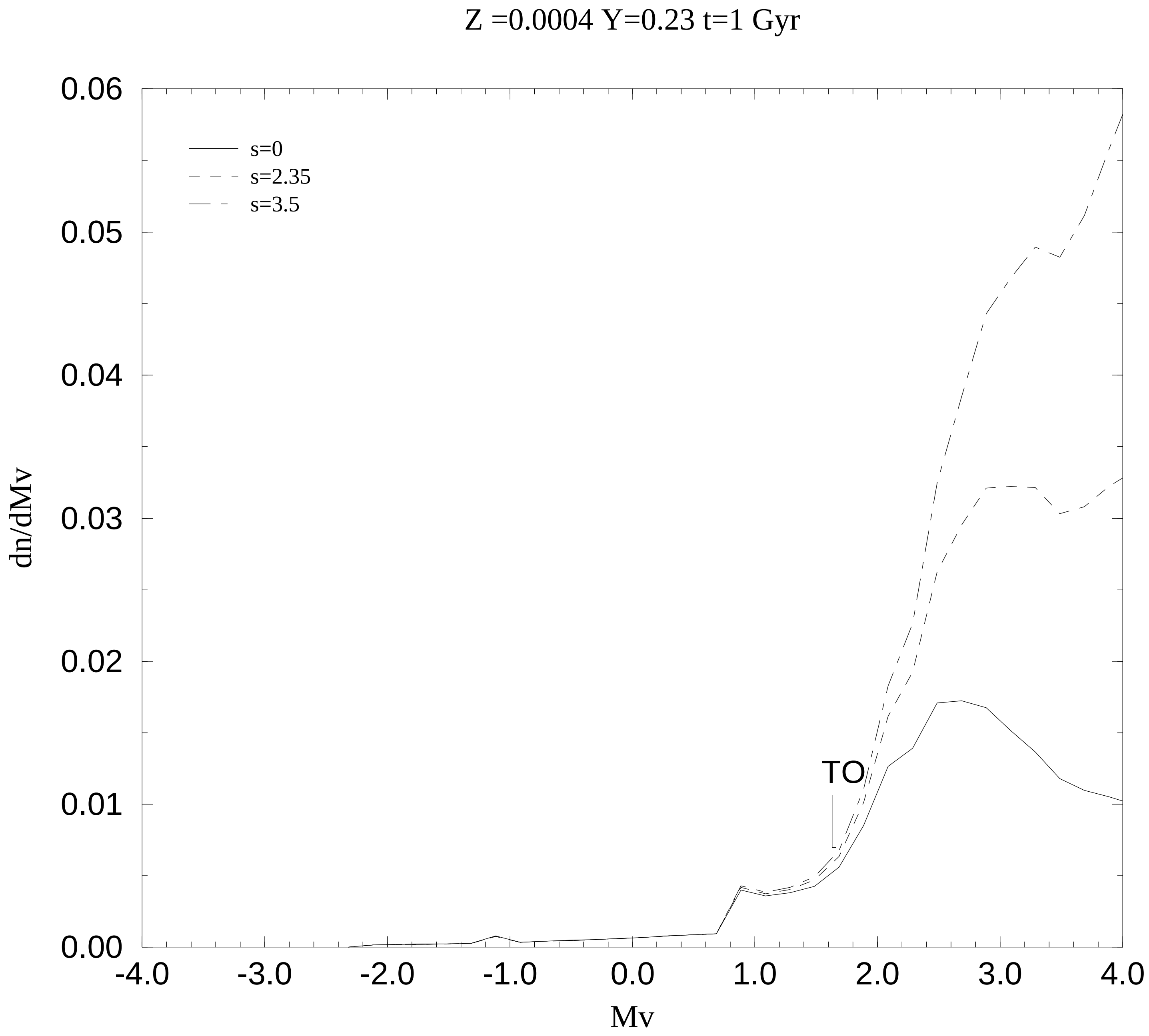

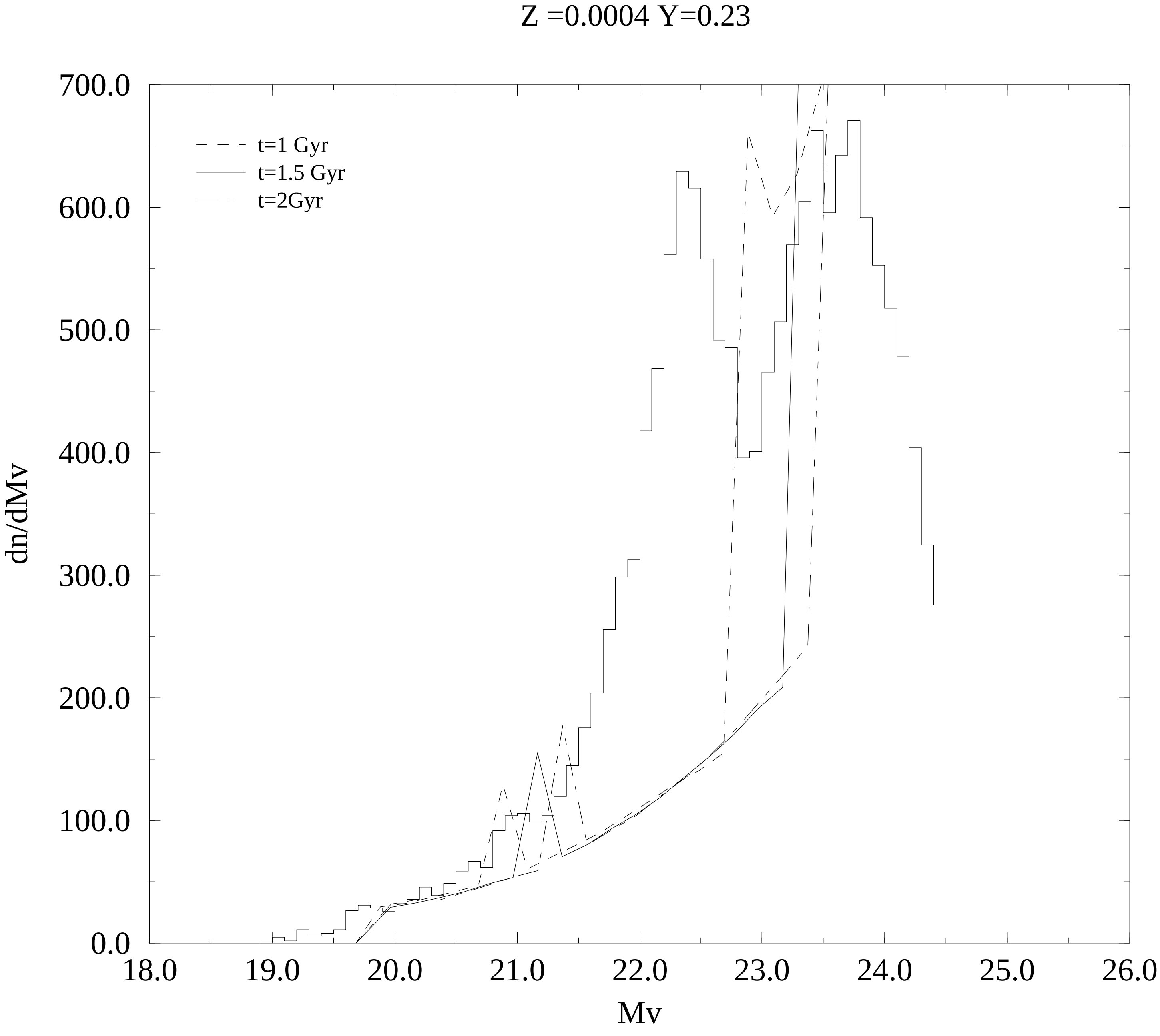